\documentstyle[12pt]{article}
\input amssym.def

\font\twelvemsa=msam10 scaled \magstep1
\font\ninemsa=msam9

\textfont\msafam=\twelvemsa
\scriptfont\msafam=\ninemsa
\font\twelvemsb=msbm10 scaled \magstep1
\font\ninemsb=msbm9

\textfont\msbfam=\twelvemsb
\scriptfont\msbfam=\ninemsb
\font\twelveeufm=eufm10 scaled \magstep1
\font\nineeufm=eufm9

\textfont\eufmfam=\twelveeufm
\scriptfont\eufmfam=\nineeufm
\newcommand{\D}{{\cal D}}
\newcommand{\gl}{{\widetilde{gl}}}
\newcommand{\hd}{{\widehat{\cal D}}}
\newcommand{\hf}{{\frac{1}{2}}}
\newcommand{\hgl}{{\widehat{gl}}}
\newcommand{\W}{{\cal W}}
\newcommand{\Wi}{{\W_{1+\infty}}}

\newcommand{\Res}{\mathop{\rm Res}\nolimits}

\newcommand{\wt}{{\mathop{\rm wt}\nolimits}}
\newcommand{\ch}{{\mathop{\rm ch}\nolimits}}
\newcommand{\tr}{{\mathop{\rm tr}\nolimits}}
\newcommand{\qed}{\hfill$\Box$}

\newtheorem{theorem}{Theorem}[section]
\newtheorem{definition}{Definition}[section]
\newtheorem{lemma}{Lemma}[section]
\newtheorem{example}{Example}[section]

\newcounter{remark}[section]
\renewcommand{\theremark}{\arabic{section}.\arabic{remark}}
\newenvironment{remark}%
{\refstepcounter{remark}\trivlist \item[\hskip
    \labelsep {\bf Remark \theremark}]}%
{\endtrivlist}

\newtheorem{corollary}{Corollary}[section]
\newtheorem{conjecture}{Conjecture}[section]
\newtheorem{proposition}{Proposition}[section]

\begin{document}
\begin{titlepage}
\title{$\W_{1+\infty}\/$ and $\W(gl_N)\/$ with central charge
  $N\/$}

\author{Edward Frenkel${}^1$\thanks{Supported by a Junior Fellowship from
Harvard Society of Fellows and in part by NSF grant DMS-9205303.},
  Victor Kac${}^2$\thanks{Supported in part by NSF grant
    DMS-9103792.},
  Andrey Radul${}^2$
  and Weiqiang Wang${}^2$\\
 {\small ${}^1$Department of Mathematics, Harvard University,
   Cambridge 02138}\\ {\small ${}^2$Department of Mathematics,
   MIT, Cambridge 02139}}
\date{}
\end{titlepage}
\maketitle

\section*{Introduction}

The Lie algebra $\hd\/$, which is the unique non-trivial central
extension of the Lie algebra $\cal D\/$ of differential operators
on the circle [KP1], has appeared recently in various models of
two-dimensional quantum field theory and integrable systems, {\em
cf.}, {\em e.g.}, [BK, FKN, PRS, IKS, CTZ, ASvM].  A systematic
study of representation theory of the Lie algebra $\hd\/$, which
is often referred to as $\Wi\/$ algebra, was initiated in [KR]. In
that paper irreducible quasi-finite highest weight
representations of $\hd\/$ were classified and it was shown that
they can be realized in terms of irreducible highest weight
representations of the Lie algebra of infinite matrices.

In the first part of the present paper we recall some of the
results of [KR] and, as an immediate corollary, obtain complete
and specialized character formulas for an arbitrary {\em
primitive} representation of $\hd\/$. (A primitive
representation of $\hd\/$ is an ``analytic continuation'' of a
quasi-finite irreducible unitary representation of $\hd\/$.) The
results of [KR] were used previously in [Mat,AFOQ] to derive
character formulas of primitive representations of central charge
$c=1\/$.

In the second part of the paper we exhibit a connection between $\hd\/$ and
the $\W\/$--algebra $\W(gl_N)\/$ at the central charge $N\/$.  Our main
result is that any primitive representation of $\hd\/$ with central charge
$N\/$ has a canonical structure of an irreducible representation of
$\W(gl_N)\/$ with the same central charge and that all irreducible
representations of $\W(gl_N)\/$ with central charge $N\/$ arise in this
way. An immediate corollary is a character formula for these
representations.

The vacuum module of $\hd\/$ of central charge $c\/$ is
irreducible if and only if $c\/$ is non-integral [KR]. If $c\/$
is a positive integer $N\/$, then this vacuum module contains a
unique singular vector of degree $N+1\/$, and the quotient by the
submodule generated from this singular vector is an irreducible
$\hd\/$-module [KR]. We will show that this quotient is
isomorphic to the vacuum module of the $\W\/$--algebra
$\W(gl_N)\/$ with the same central charge. All these modules
carry vertex operator algebra (or chiral algebra) structures, and
this isomorphism holds at the level of vertex operator algebras.
It follows that the Lie algebra of Fourier components of the
fields from $\W(gl_N)\/$ with central charge $N\/$ is a quotient
of the local completion $U_N(\hd)_{loc}\/$ of the universal
enveloping algebra of $\hd\/$ with the same central charge, by a
certain ideal. A similar statement was conjectured in [FKN],
where it was used in the study of the so-called
$\W\/$-constraints in two-dimensional quantum gravity.

We recall that the $\W\/$--algebra
$\W(gl_N)\/$ can be defined as the kernel of certain
``screening'' operators acting on bosonic Fock spaces, {\em cf.}
[FF2]. These operators depend on a complex parameter $\beta\/$,
and for ``generic'' values of $\beta\/$, this kernel is finitely
generated as a chiral algebra. We show that $\beta = 1\/$ is a
generic value; this value corresponds to central charge $N\/$,
{\em cf.} also [F2, Bo, BS]. We then use the realization of
$\hd\/$ in terms of $N\/$ free bosonic fields to construct an
explicit map from the chiral algebra of $\hd\/$ to the chiral
algebra $\W(gl_N)\/$, and to show that this map is an
isomorphism.
It follows that under this map of chiral algebras, the first
$N\/$ generating fields $J^0(z),\ldots,J^{N-1}(z)\/$ of $\hd\/$
map to the generating fields $W^0(z),\ldots,W^{N-1}(z)\/$ of
$\W(gl_N)\/$, and the remaining fields $J^m(z), m \geq N\/$, map
to certain normally ordered combinations of
$W^0(z),\ldots,W^{N-1}(z)\/$ and their derivatives. This happens
so because by taking the quotient by a submodule generated from
the singular vector (this is often referred to as decoupling of a
singular vector), we effectively set to zero a field of the form
$$ J^N(z) - :P(J^0(z),\ldots,J^{N-1}(z)):,
$$
where the second term is a normally ordered polynomial in
$J^0(z), \ldots, J^{N-1}(z)\/$ and their derivatives. This allows
one to express $J^m(z), m \geq N\/$, as a combination of $J^0(z),
\ldots, J^{N-1}(z)\/$ and their derivatives. Since such
combinations are non-linear, the resulting commutation relations
between $W^0(z), \ldots, W^{N-1}(z)\/$ also become non-linear,
which is what we expect in $\W(gl_N)\/$.

One could try to prove this correspondence between $\hd\/$ and $\W(gl_N)\/$
by using an explicit formula for the singular vector in the vacuum module
of $\hd\/$. Although we know a simple formula for this vector in the Verma
module over $\widehat{\cal D}\/$ [KR, Sect.~5.2], for large $N\/$ it is
difficult to derive from it a formula for such a vector in the vacuum
module in the PBW basis. But even if we knew a precise formula for it, it
would still be unclear how to show that decoupling of this vector gives
$\W(gl_N)\/$. That is why we prefer a more indirect, but simpler and more
transparent proof, which uses free field realizations of $\hd\/$ and
$\W(gl_N)\/$.

Our result implies that any irreducible representation of
$\W(gl_N)\/$ with central charge $N\/$ gives rise to a
quasi-finite irreducible representation of $\hd\/$ with the same
central charge. Irreducible representations of $\W(gl_N)\/$ can
be constructed as submodules of the Fock modules over the
Heisenberg algebra of $N\/$ scalar fields. They yield primitive
representations of $\hd\/$ and all of them can be constructed in
this way.

In the third part of the paper we establish a remarkable duality between
``integral'' irreducible representations of ${\cal W} \left(gl_N \right)\/$
and finite-dimensional irreducible representations of $GL_N (\Bbb C)\/$,
{\em cf.} also [K, F1, F2, KP1]. This leads us to the conjecture that the
fusion algebra of integral representations of ${\cal W} \left(gl_N
\right)\/$ is isomorphic to the representation algebra of the group $GL_N
(\Bbb C)\/$.

The paper is organized as follows. In Sect.~1 and the first part
of Sect.~2 we set notation and recall some of the results of
[KR]. In the second part of Sect.~2 we establish character
formulas for primitive representations of $\hd\/$ with central
charge $N\/$. In Sections 3 and 4 we study the vertex operator
algebra structure on the vacuum module of $\hd\/$ and on
$\W(gl_N)\/$. In Sect.~5 we construct a surjective homomorphism
between them and derive consequences of this fact.  In Sect.~6 we
establish the duality between ${\cal W} \left(gl_N \right)\/$ and
$GL_N (\Bbb C)\/$.

\section{The Lie algebra $\hd\/$}

Let $\cal D \/$ be the Lie algebra of complex regular
differential operators on ${\Bbb C}^{\times}\/$ with the usual
bracket, in a indeterminate $t\/$. The elements
\begin{equation}
  J^l_k = - t^{l+k} (\partial_t)^l,\quad (k \in \Bbb Z, l \in
  \Bbb Z_{+})
\label{eq:one}
\end{equation}
form a basis of $\cal D\/$. The Lie algebra $\cal D\/$ has the
following 2-cocycle with values in $\Bbb C\/$ [KP1, p.3310]:

\begin{equation}
  \Psi (f(t)(\partial_t)^m, g(t)(\partial_t)^n) =
  \frac{m!n!}{(m+n+1)!} \Res_{t=0} f^{(n+1)}(t) g^{(m)}(t)dt,
\label{eq:cocy}
\end{equation}
where $f^{(m)}(t) = \partial_t^m f(t)\/$. We denote by $\hd=
{\cal D} \oplus {\Bbb C} C\/$, where $C\/$ is the central
element, the corresponding central extension of the Lie algebra
$\cal D\/$.

Another important basis of $\cal D\/$ is
\begin{equation}
  L^l_k = - t^{k} D^l \quad (k \in \Bbb Z, l\in \Bbb Z_{+})
\label{eq:two}
\end{equation}
where $D = t\partial_t\/$. These two bases are
related by the formula [KR]:
\begin{equation}
  J^l_k = -t^k [D]_l.
\label{eq:3}
\end{equation}
Here and further we use the usual notation $[x]_l = x (x-1)\ldots
(x-l+1)\/$. One has another formula for this cocycle [KR]
\begin{equation}
  \Psi \left( t^r f(D), t^s g(D) \right)= \left\{
  \begin{array}{ll}
    \sum_{-r \leq j \leq -1} f(j) g(j+r), & \mbox{if } r=-s \geq
    0,\\ 0, & \mbox{if } r+s \neq 0.
  \end{array} \right.
\label{eq:4}
\end{equation}
The bracket in $\hd\/$ may be conveniently calculated by the
following formula:
\begin{equation}
  [t^r f(D), t^s g(D)] = t^{r+s} \left( f(D+s)g(D)-f(D)g(D+r)
  \right) + \Psi (t^r f(D), t^s g(D)) C.
\label{eq:bracket}
\end{equation}

The Lie algebra $\hd\/$ contains two 1-parameter families of
Virasoro subalgebras $Vir^{\pm}(\beta)\/$, $\beta \in \Bbb C\/$,
defined by
\begin{eqnarray}
  L_k^{+}(\beta) &=& L_k^1 + \beta (k+1) L_k^0, \label{eq:vira}\\
  L_k^{-}(\beta) &=& L_k^1 + (k+\beta (-k+1)) L_k^0, \nonumber
\end{eqnarray}
so that
\begin{equation}
  [L_m^{\pm}(\beta), L_m^{\pm}(\beta)]
  = (m-n) L_{m+n}^{\pm}(\beta)
  + \delta_{m,-n} \frac{m^3 -m}{12} C_{\beta},
\label{eq:family}
\end{equation}
where $C_{\beta}= -(12 \beta^2 - 12 \beta +2)C\/$\footnote{The
value of $C_{\beta}\/$ given in [KR] should be corrected.}.  Note
that these two families intersect at $\beta = \frac{1}{2}\/$ and
that $C_{\frac{1}{2}} = C\/$.

As in [KR], define an anti-linear anti-involution $\sigma\/$ of
$\hd\/$ by:
\begin{equation}
  \sigma(t^k f(D)) = t^{-k} \bar{f} (D-k),\quad
  \sigma (C) = C,
\label{eq:5}
\end{equation}
where for $f(D)=\sum_i f_i D^i, f_i \in \Bbb C\/$, we let
$\bar{f}(D) = \sum_i \bar{f}_i D^i\/$.  Then we have $\sigma
L_k^{+}(\beta) = L_{-k}^{-}(\beta)\/$.  In particular,
$Vir(\frac{1}{2}):=Vir^{\pm}(\frac{1}{2})\/$ is the only $\sigma\/$-stable
subalgebra among the $Vir^{\pm}(\beta)\/$.

Define a $\Bbb Z\/$-gradation $\hd = \oplus_{j \in \Bbb Z}
\widehat{\cal D}_j \/$ by letting
$$
  \wt  L_k^l = \wt J_k^l = k, \quad \wt C =0.
$$
This gives us the triangular decomposition of $\hd\/$:
\begin{equation}
  \hd = \widehat{\cal D}_{+} \oplus\widehat{\cal D}_0
  \oplus \widehat{\cal D}_{-},
\label{eq:6}
\end{equation}
where $\widehat{\cal D}_{\pm} = \oplus_{j \in \pm \Bbb N}
\widehat{\cal D}_j\/$, $\widehat{\cal D}_0 = {\cal D}_0 \oplus
\Bbb C C\/$.  Note that $\sigma (\hd_j ) = \hd_{-j}, \sigma
(\hd_{+} ) = \hd_{-}\/$,\linebreak $\sigma (\hd_0) = \hd_0\/$.

Fix $c \in \Bbb C\/$. Given $\lambda \in {\cal D}_0^{*}\/$, we
define in a standard way the {\em Verma module\/} with central
charge $c\/$ over $\widehat{\cal D}\/$:
$$
  M_c (\widehat{\cal D}, \lambda) = U (\widehat{\cal D})
  \otimes_{U (\widehat{\cal D}_0 \oplus \widehat{\cal D}_{+})}
  \Bbb C_{\lambda},
$$
where $\Bbb C_{\lambda}\/$ is the 1-dimensional $\widehat{\cal
D}_0 \oplus \widehat{\cal D}_{+}\/$-module, on which $C\/$ acts
as multiplication by $c\/$, $h \in \widehat{\cal D}_0\/$ acts as
multiplication by $\lambda(h)\/$, and $\widehat{\cal D}_{+}\/$
acts by $0\/$. Here and further we denote by $U (\frak g)\/$ the
universal enveloping algebra of a Lie algebra $\frak g\/$.  In
general, we shall say that a $\hd\/$-module has central charge $c
\in \Bbb C\/$ if $C\/$ acts on it by multiplication by $c\/$.

Denote by $\cal P\/$ the subalgebra of $\cal D\/$, which consists
of the operators that can be extended to regular differential
operators on $\Bbb C\/$. We have:
$$
  {\cal P} = \mbox{linear span of }
  \left\{ J^l_k| l+k\geq 0 \right\}.
$$
It follows from (\ref{eq:cocy}) that $\cal P\/$ is a subalgebra of
$\hd\/$.  Let $\widehat {\cal P} = {\cal P} \oplus \Bbb C C\/$.
Note that $\widehat{\cal D}_0 \oplus \widehat{\cal D}_{+}
\subset \widehat{\cal P}\/$ and that the $\widehat{\cal D}_0
\oplus \widehat{\cal D}_{+}\/$-module $\Bbb C_0\/$ can be
extended to be a $\widehat{\cal P}\/$-module by
letting ${\cal P} \mapsto 0\/$. The induced $\hd\/$-module
$$
  M_c(\hd; \widehat {\cal P})
  = U (\hd)\otimes_{U (\widehat {\cal P})} \Bbb C_0,
$$
which is a quotient of the Verma module $M_c(\hd, 0)\/$,
is called the {\em vacuum $\hd\/$-module\/} with central charge
$c\/$.

There exists a unique irreducible quotient of the Verma module
$M_c (\hd, \lambda)\/$, denoted by $V_c (\hd, \lambda)\/$.
The module $V_c (\hd, \lambda)\/$ is called {\em
quasi-finite} if all eigenspaces of $D\/$ are finite-dimensional
(note that $D\/$ is diagonalizable). It was proved in [KR, Theorem 4.2] that
$V_c (\hd, \lambda)\/$ is a quasi-finite module if and only if
the generating series
$$
  \Delta_{\lambda} (x)
  = \sum_{n=0}^{\infty} \frac{x^n}{n!} \lambda \left(L^n_0\right)
$$
has the form:
\begin{equation}
  \Delta_{\lambda} (x) = \frac{\phi (x)}{e^x -1},
\label{eq:7}
\end{equation}
where $\phi (x)\/$ is a quasi-polynomial ({\em i.e.} a linear
combination of functions of the form $x^n e^{\alpha x}\/$, where
$n \in \Bbb Z_{+}\/$ and $\alpha \in \Bbb C \/$) such that $\phi
(0) = 0\/$.

Furthermore, it was shown in [KR, Theorem~5.2] that $V_c (\hd,
\lambda)\/$ is a non-trivial unitary module with respect to the
anti-involution $\sigma\/$ if and only if $c\/$ is a positive
integer and
\begin{equation}
  \Delta_{\lambda} (x) = \sum_{i=1}^{c}  \frac{e^{r_i x}-1}{e^x-1}
\label{eq:delta}
\end{equation}
for some $r_1, \ldots\/$, $r_c \in \Bbb R\/$.

\begin{definition} \label{def:1}
The $\hd\/$-module $V_c (\hd, \lambda)$ with c a positive integer
and $\Delta_{\lambda}(x)$ of the form (\ref{eq:delta}) with $r_1,
\ldots$, $r_c \in \Bbb C$ is called a {\em primitive}
$\hd\/$-module.  The numbers $r_1, \ldots, r_c$ are called the
\/{\em exponents} of this module.
\end{definition}

\section{Characters of primitive $\hd\/$-modules}
\setcounter{equation}{0}

Let $\gl\/$ be the Lie algebra of all matrices $(a_{ij})_{i,j \in
{\Bbb Z}}\/$ with only finitely many nonzero diagonals.  Letting
$\wt E_{ij} = j-i\/$ defines a $\Bbb Z\/$-gradation $\gl =
\oplus_{j \in \Bbb Z}\gl_{j}\/$.  Given $s \in \Bbb C\/$, we may
consider the natural action of $\hd\/$ on the space $t^s \Bbb C
[t, t^{-1}]\/$. Choosing the basis $v_j = t^{-j+s}\/$ ($j \in
\Bbb Z\/$) of this space gives us a homomorphism of Lie
algebras $\phi_{s}: {\cal D} \rightarrow \gl\/$:
\begin{equation}
  \phi_{s} \left(t^k f(D) \right) = \sum_{j \in {\Bbb Z}} f(-j+s)
  E_{j-k,j}.
\label{eq:fis}
\end{equation}
Denote by $\hgl =\gl \oplus \Bbb C K\/$ the central
extension given by the 2-cocycle [KP1]
\begin{displaymath}
  C(A,B) = \tr
([J,A]B), \mbox{ where } J=\sum_{i \leq 0} E_{ii}.
\end{displaymath}
The $\Bbb Z\/$-gradation of $\gl\/$ extends to $\hgl\/$ by
letting $\wt K = 0\/$. The Lie algebra $\hgl\/$ has the
following antilinear anti-involution:
$$
  A \longmapsto {}^t \bar{A}, \quad
  K \longmapsto K,
$$
where ${}^t \bar{A}\/$ stands for the hermitean conjugate
of a matrix $A\/$.

The map $ \widehat{\phi}_s: \widehat{\cal D} \mapsto \hgl\/$
defined by
\begin{eqnarray}
  \widehat{\phi}_s \mid_{{\widehat{\cal D}}_j}
  & = & \phi_s\mid_{{\widehat{\cal D}}_j} \mbox{ if } j \neq 0,
  \label{eq:13} \\
  \widehat{\phi}_s (e^{xD}) & = & \phi_s (e^{xD}) -
  \frac{e^{sx}-1}{e^{x} -1} K, \quad \widehat{\phi}_s (C) = K
  \nonumber
\end{eqnarray}
is an injective homomorphism compatible with the $\Bbb Z\/$-gradations and
the involutions [KR]. Let $\hgl^m \/$ be the direct sum of $m\/$ copies of
$\hgl\/$.  Given ${\bf s} = (s_1, \ldots, s_m) \in \Bbb C^m\/$ we have a
homomorphism $\widehat{\phi}_{\bf s} = \oplus_i
\widehat{\phi}_{s_i}: \widehat{\cal D} \mapsto \hgl^m\/$.

Given $\lambda \in \gl_0^*\/$ and $c \in \Bbb C\/$, there exists
a unique irreducible $\hgl\/$-module $V_c(\hgl, \lambda)\/$ with
$K = cI\/$, which admits a non-zero vector $|\lambda \rangle\/$
such that
\begin{eqnarray*}
  E_{ij} |\lambda \rangle &= &0\quad \mbox{for } i<j,\\
  E_{ii} |\lambda \rangle &= &\lambda (E_{ii})|\lambda \rangle
  \quad \mbox{for } i \in \Bbb Z.
\end{eqnarray*}
All the modules $V_c(\hgl, \lambda)\/$ are quasi-finite in the
sense that all the eigenspaces of $\widehat{\phi}_0 (D)\/$ (and
hence of $\widehat{\phi}_s (D)\/$, $s \in \Bbb C\/$) are
finite-dimensional.

Define $\Lambda_j \in \gl_0^*\/$ ($j \in \Bbb Z\/$) as follows:
\begin{equation}
  \Lambda_j (E_{ii}) =
  \left\{
  \begin{array}{rl}
    1  & \mbox{for } 0<i \leq j, \\
    -1 & \mbox{for } j<i \leq 0, \\
    0  & \mbox{otherwise.}
  \end{array} \right.
\label{eq:14}
\end{equation}
Then a $\hgl\/$-module $V_c(\hgl, \lambda)\/$ is a non-trivial
unitary module if and only if $c\/$ is a positive integer and
\begin{equation}    \label{lam}
  \lambda = \Lambda_{n_1}+\Lambda_{n_2}+\ldots +\Lambda_{n_c},
  \mbox{ where } n_1\geq n_2\geq \ldots \geq n_c.
\end{equation}

One has the following ``specialized'' character formula for these
modules [KP2]:
\begin{equation}    \label{eq:char}
  \tr_{V_c(\widehat{gl},\lambda)} q^{\widehat{\phi}_s (L_0^1)}
  = q^{a(\lambda)} \prod_{1 \leq i < j \leq c}
  (1-q^{n_i-n_j+j-i}) / \varphi (q)^{c}
\end{equation}
where $\lambda\/$ is given by (\ref{lam}), $a(\lambda) = \sum_k
(n_k+s)(n_k+s+1)/2\/$, and $\varphi (q) =
\prod_{j=1}^{\infty} \left( 1-q^j \right)\/$ is the Euler
product.

It is proved in [KR, Theorem~4.5] that an irreducible
quasi-finite $\hgl^m\/$-module remains irreducible when
restricted to $\widehat{\phi}_{\bf s} (\hd)\/$, provided that
$s_i - s_j \not \in \Bbb Z\/$ for $i\neq j\/$.  This allows one
to describe the primitive $\hd\/$-modules.

\begin{proposition}[KR, Theorem 4.6]
Let $V$ be a primitive $\hd$-module with exponents $r_1, \ldots,
r_c$. Break the set $\{r_1, \ldots, r_c \}$ into a disjoint union
of congruent mod $\Bbb Z$ classes, {\em i.e.}
$$
\{ r_1, \ldots, r_c \}
= S_1 \bigcup \ldots \bigcup S_m,
$$
where $S_i = \left\{s_i + n_1^{(i)}, \ldots, s_i + n_{c_i}^{(i)}
\right\}$, $n_j^{(i)} \in \Bbb Z$ and $s_i - s_j \notin \Bbb Z$.  Let
${\bf s} = (s_1, \ldots, s_m)$ and $\Lambda^{(i)}(V) =
\Lambda_{n_1^{(i)}} + \ldots + \Lambda_{n_{c_i}^{(i)}}$.  Then
the $\hd$-module $V$ is obtained from the $\hgl^m$-module
$$
  \bigotimes_{i=1}^m V_{c_i} \left( \hgl, \Lambda^{(i)}(V) \right)
$$
by restricting to $\widehat{\phi}_{\bf s}(\hd)$.  In particular
the specialized character $tr_V q^{L^1_0}$ is equal to the
product of the corresponding characters of the irreducible
$\hgl$-modules (given by the right-hand side of (\ref{eq:char})).
\label{prop_ch}
\end{proposition}

Let $H_i = E_{i,i} - E_{i+1,i+1} + \delta_{i0} K\/$ ($i \in
\Bbb Z\/$) be the simple coroots of $\hgl\/$. We define
$\widehat{\Lambda}_0 \in \hgl_0^*\/$ by $\widehat{\Lambda}_0 (K)
= 1\/$, $\widehat{\Lambda}_0 (E_{ii}) = 0\/$ for all $i\/$ and
extend $\Lambda_j\/$ from $\gl_0^*\/$ to $\hgl_0^*\/$ by letting
$\Lambda_j (K) =0\/$. Then $\widehat{\Lambda}_j = \Lambda_j +
\widehat{\Lambda}_0\/$ ($j \in \Bbb Z\/$) become the fundamental
weights, {\em i.e.} $\widehat{\Lambda}_j (H_i)= \delta_{ij}\/$.

The highest weight of a unitary module $V_c(\hgl, \lambda)\/$ is
defined as $\widehat{\lambda} = \lambda + c \widehat{\Lambda}_0\/$.
We have
\begin{equation}
  \widehat{\lambda} = \sum_i {k_i \widehat{\Lambda}_i}, \quad
  \mbox{where} \quad
  k_i \in \Bbb Z_{+} \mbox{ and } \sum_i k_i = c.
\label{eq:8}
\end{equation}
We shall often write $V (\widehat{gl}, \widehat\lambda)\/$ in place of
$V_c (\widehat{gl}, \lambda)\/$.

Let $p(\widehat{\lambda})= \{p_i\}\/$ be the sequence associated to
$\widehat{\lambda}\/$ which is non-increasing and contains $k_j\/$ integers
equal $j\/$ ($j \in \Bbb Z\/$).  Because of the obvious stabilization
properties, the classical character formula for $gl(N, \Bbb C)\/$, {\em
cf.} [M, Ch.~1, (2.9$'$)], still holds for $\hgl\/$ and can be stated as
follows. Define the complete character of the module $V_c(\hgl, \lambda)\/$
by
$$
\ch_{c,\lambda} (h) (= \ch_{\widehat\lambda} (h)) = \tr_{V_c(\widehat{gl},
\lambda)} e^h, \quad
h \in \hgl_0.
$$
Let $ S_n = \ch_{\widehat\Lambda_n}\/$ ($n \in \Bbb Z\/$) be the
character of the $n\/$-th fundamental module. Then we have:
\begin{equation}
  \ch_{c,\lambda} = \mbox{det} (S_{p_i-i+j})_{i,j= 1, \ldots, c.}
\label{eqn_det}
\end{equation}

\noindent{\em Example.}
$$
\ch_{2 \widehat\Lambda_n} = S_n^2  - S_{n-1}S_{n+1}.
$$

We can now define the {\em complete character\/} of a
$\hd\/$-module $V\/$ by
$$
  \ch V = \tr_V \prod_{n=0}^{\infty} x_n ^{L_0^n}.
$$
Due to Proposition \ref{prop_ch} and formula (\ref{eqn_det}), the
calculation of the characters of primitive $\hd\/$-modules
reduces to the calculation of characters of fundamental
$\hgl\/$-modules restricted to $\widehat{\phi}_s (\hd)\/$.

Let ${\cal F} = \oplus_{i \in \Bbb Z} V (\hgl,
\widehat{\Lambda}_i)\/$ denote the direct sum of all the
fundamental $\hgl\/$-modules.  The following construction of
$\cal F\/$ is well known (see {\em e.g.}, [K2, Chapter~14]).
Fix $s \in \Bbb C\/$ and consider the Clifford algebra $Cl\/$
over $\Bbb C\/$ on generators $\psi_j\/$ and $\psi_j^*\/$ ($j
\in \Bbb Z\/$) with defining relations
\begin{equation}
  \left[ \psi_i, \psi^*_j \right]_+ = \delta_{i, -j}, \quad
  \left[\psi_i, \psi_j \right]_+ = 0, \quad
  \left[\psi^*_i, \psi^*_j \right]_+ = 0.
\label{eq:9}
\end{equation}
Then $\cal F\/$ is identified with the space of the unique
irreducible representation of the algebra $Cl\/$ which admits a
non-zero vacuum vector $| 0 \rangle\/$ such that
\begin{displaymath}
  \psi_j | 0 \rangle = 0 \mbox{ if } j \geq 0, \quad
  \psi^*_j | 0 \rangle = 0 \mbox{ if } j > 0.
\end{displaymath}
The basis element $E_{ij}\/$ of $\gl\/$ is represented by the
operator $: \psi_{-i} \psi^*_j :\/$ ($= \psi_{-i} \psi^*_j\/$ if
$j > 0\/$ and $= - \psi^*_j \psi_{-i}\/$ otherwise) and $K\/$
by the identity operator.  The decomposition of $\cal F\/$ into
irreducibles with respect to $\widehat{gl}\/$ coincides with the charge
decomposition ${\cal F} = \bigoplus_{m \in {\Bbb Z}} {\cal
  F}^{(m)}\/$, where charge $| 0 \rangle = 0\/$, charge $\psi_j =
-\/$ charge $\psi^*_j = 1\/$.  Due to (\ref{eq:fis}), we have
\begin{displaymath}
  \left[\widehat\phi_s (L_0), \psi_r \right] = - (r + s) \psi_r,
  \quad
  \left[\widehat\phi_s (L_0), \psi^*_r \right] = (- r + s) \psi^*_r .
\end{displaymath}
Since the vectors $\psi_{-i_1} \ldots \psi_{-i_a} \psi_{-j_1}^{*}
\ldots \psi_{-j_b}^{*} |0\rangle\/$ with $0< i_1 < i_2 <
\ldots\/$, and $0 \leq j_1 < j_2 < \ldots\/$, form a basis of
$\cal F\/$, we obtain the following formula for $\ch {\cal F}\/$
restricted to $\widehat{\phi}_s (\hd)\/$ (cf.\ [AFOQ]):
\begin{equation}
  \ch {\cal F} = \prod_{r\in \Bbb Z_{+}}
  (1+ \prod_{n\in \Bbb Z_{+}} x_n^{-(r+1+s)^n})
  (1+ \prod_{n\in \Bbb Z_{+}} x_n^{(-r+s)^n})
\label{eq:10}
\end{equation}
Then the character of $m\/$-th fundamental $\hgl\/$-module
restricted to $\widehat{\phi}_s (\hd)\/$ is equal to
\begin{equation}
  S_m (x) = x_0^{-m}\Res_{x_0 = 0} x_0^{m-1} \ch {\cal F} dx_0.
\label{eq:jap}
\end{equation}

Summarizing, we obtain the following result:

\begin{theorem}    \label{th:sumch}
  Let $V$ be a primitive $\hd$-module with exponents $r_1,
  \ldots, r_c$. We keep notation of Proposition \ref{prop_ch}.
  Let $\widehat{\Lambda}^{(k)}(V) = {\Lambda}^{(k)}(V) + c_k
  \widehat{\Lambda}_0\/$, $k = 1, \ldots , m$. Then the complete
  character of $V$ is given by
  $$ \ch V = \prod_{k=1}^m \ch_{\widehat{\Lambda}^{(k)}(V)} (x),
  $$
  where $\ch_{\widehat{\Lambda}}(x) = \mbox{det} \left(S_{p_i -i +
    j} (x)\right)_{i,j = 1, \ldots,c}\/$, $\{ p_i \} =
  p(\widehat{\Lambda})$    and $S_n (x)$ ($n \in \Bbb
  Z$) are given by formulas (\ref{eq:10}) and (\ref{eq:jap}).
\end{theorem}

\section {VOA structure on the vacuum module of $\hd\/$.}
\setcounter{equation}{0}

In this section we will define the structure of a vertex operator
algebra (VOA) on the vacuum module $M_c\/$ and hence on the
irreducible quotient module $V_c\/$ over $\hd\/$. The general
definition of VOA was given in [B, FLM]. We will however use a
slightly different approach, inspired by [G]. This approach will
allow us to give a simple proof that this structure indeed
satisfies all axioms of VOA.

Let $V=\oplus_{n=0}^{\infty}V_n\/$ be a $\Bbb Z_{+}\/$-graded
vector space, where $\dim V_n < \infty\/$ for all $n\/$, called
the {\em space of states\/}. A {\em field\/} on $V\/$ of
conformal dimension $\Delta \in \Bbb Z\/$ is a power series $\phi
(z) = \sum_{j \in \Bbb Z} \phi_j z^{-j - \Delta}\/$, where
$\phi_j \in \mbox{End}\, V \/$ and $\phi_j V_n \subset
V_{n-j}\/$. Note that if $\phi(z)\/$ is a field of conformal
dimension $\Delta\/$, then the power series $\partial_z \phi (z)
= \sum_{j \in \Bbb Z} (-j - \Delta)\phi_j z^{-j - \Delta-1}\/$ is
a field of conformal dimension $\Delta +1\/$. Let
\begin{equation}
  \phi_{+}(z) = \sum_{j > - \Delta}\phi_j z^{-j - \Delta}, \quad
  \phi_{-}(z) = \sum_{j \leq - \Delta}\phi_j z^{-j - \Delta}.
\label{eq:11}
\end{equation}

Given two fields $\phi (z)\/$ and $\psi(z)\/$ of conformal dimensions
$\Delta_\phi$ and $\Delta_\psi$ one defines their {\em normally ordered
product\/} as the field
\begin{equation}
  :\phi (z)\psi(z): = \phi_{-}(z)\psi(z)+\psi(z)\phi_{+}(z)
\label{eq:normal}
\end{equation}
of conformal dimension $\Delta_\phi+\Delta_\psi$.
The Leibniz rule holds for the normally ordered product:
\begin{equation}
  \partial_z :\phi (z)\psi(z): =
  :\partial_z \phi (z)\psi(z): +:\phi (z)\partial_z \psi(z):.
\label{eq:rule}
\end{equation}

Two fields $\phi (z)\/$ and $\psi(z)\/$ are called {\em local\/}
with respect to each other, if for any $v \in V_n\/$ and $v^* \in
V_m^*\/$ both matrix coefficients $\langle v^* |\phi
(z) \psi(w) | v \rangle\/$ for $|z| > |w|\/$ and $\langle v^*
|\psi(w)\phi (z)|v\rangle\/$ for $|z| < |w|\/$ converge to the
same rational function in $z\/$ and $w\/$ which has no poles
outside the hyperplanes $z=0\/$, $w=0\/$ and $z=w\/$.

A {\em VOA structure\/} on $V\/$ is a linear map (the state-field
correspondence) $Y(\cdot,z) : V \longrightarrow \mbox{End}\, V
\left[\left[z,z^{-1}\right]\right]\/$ which associates to each $a
\in V_n\/$ a field of conformal dimension $n\/$ (also called a
{\em vertex operator\/}) $Y(a,z) = \sum_{j \in \Bbb Z}a_j
z^{-j-n}\/$, such that the following axioms hold:

{\bf (A1)} (vacuum axiom) There exists an element $|0\rangle \in
V_0\/$ such that $Y(|0\rangle ,z) = \mbox{Id}\/$ and $\lim_{z
\rightarrow 0} Y(a,z)|0\rangle = a\/$.

{\bf (A2)} (translation invariance) There exists an operator $T
\in \mbox{End}\, V\/$ such that
$$
\partial_z Y(a,z) = Y(T a,z) = [T, Y(a,z)].
$$

{\bf (A3)} (locality) All fields $Y(a,z)\/$ are local with respect to each
other.

A VOA $V\/$ is called {\em conformal\/} of {\em rank} $c \in \Bbb
C\/$ if there exists an element $\omega \in V_2\/$ (called the
{\em Virasoro element\/}), such that the corresponding vertex
operator $Y(\omega,z)=\sum_{n\in \Bbb Z} L_n z^{-n-2}\/$
satisfies the following properties:

{\bf (C)} $L_{-1} = T\/$, $L_0 |_{V_n} = n \cdot \mbox{Id}\/$, and
$L_2 \omega = \frac 12 c | 0 \rangle\/$.

One can show that a VOA automatically satisfies the associativity property:
\begin{equation}
  Y(a,z) Y(b,w) = Y ( Y(a, z-w) b, w).
\label{eq:assoc}
\end{equation}
Here the left-hand (resp.\ right-hand) side is the analytic
continuation from the domain $\left|z \right| > \left|w
\right|\/$ (resp.\ $ |w| > |z-w|\/$). Formula (\ref{eq:assoc})
gives the operator product expansions.  In particular, one easily derives
from {\bf (C)} that the $L_n\/$ form a Virasoro algebra with central charge
$c\/$.

Let us call two fields $\phi(z)$ and $\psi(z)$ {\em ultralocal} with
respect to each other if there exists an integer $N$, such that for any $v
\in V_n$ and $v^* \in V_m^*$, both series $\langle v^* |\phi (z)
\psi(w) | v \rangle\/ (z-w)^N$ and $\langle v^* |\psi (w)
\phi(z) | v \rangle\/ (z-w)^N$ are equal to the same finite polynomial in
$z^{\pm 1}$ and $w^{\pm 1}$. Clearly, ultralocality implies
locality. Moreover, in a vertex operator algebra any two vertex operators
are automatically ultralocal with respect to each other according to
formula (\ref{eq:assoc}) and the fact that the $\Bbb Z$--gradation on $V$
is bounded from below.

The following proposition allows one to check easily the axioms of a VOA.

\begin{proposition} \label{prop:generator}
  Let $V$ be a $\Bbb Z_{+}$-graded vector space.  Suppose that to some
  vectors $a^{(0)} = |0\rangle \in V_0, a^{(1)} \in V_{\Delta_1}, \ldots$,
  one associates fields $Y(|0\rangle ,z) = \mbox{Id}, Y(a^{(1)}, z) =
  \sum_j a_j^{(1)} z^{-j-\Delta_1} , \ldots $ of conformal dimensions $0,
  \Delta_1, \ldots$, such that the following properties hold:

  {\bf (1)} all fields $Y(a^{(i)}, z)$ are ultralocal with respect to
  each other;

  {\bf (2)} $ \lim_{z\rightarrow 0} Y(a^{(i)}, z)|0\rangle
  = a^{(i)}$;

  {\bf (3)} the space $V$ is spanned by the vectors
  \begin{equation}
    a^{(k_s)}_{-j_s - \Delta_{k_s}} \ldots a^{(k_1)}_{-j_1 -
      \Delta_{k_1}}|0\rangle ,
      \,\,\, j_1, \ldots, j_s \in \Bbb Z_{+};
    \label{eq:span}
  \end{equation}

  {\bf(4)} there exists an endomorphism $T\/$ of $V$ such that
  \begin{equation}
    \left[T, a^{(k)}_{-j-\Delta_k} \right] =
      (j+1)a^{(k)}_{-j-\Delta_k - 1},
    \quad
    T(|0\rangle ) =0.
    \label{eq:eq}
  \end{equation}
  Then letting
  \begin{equation}
    \begin{array}{rcl}
      \lefteqn{Y(a^{(k_s)}_{-j_s - \Delta_{k_s}} \ldots
        a^{(k_1)}_{-j_1 - \Delta_{k_1}} |0\rangle , z)} \\
      & = & (j_1! \cdot \ldots \cdot j_s!)^{-1} \cdot
      :\partial_z^{j_s}Y(a^{(k_s)},z) \ldots \partial_z^{j_2}
      Y(a^{(k_2)},z) \partial_z^{j_1} Y(a^{(k_1)},z):
    \end{array}
  \label{eq:norm}
  \end{equation}
  (where the normal ordering of more than two fields is from right to
  left as usual), gives a well-defined VOA structure on $V$.

\end{proposition}

{\em Proof\/}.  Choose a basis of monomials (\ref{eq:span}) and
construct the map $Y (\cdot, z)\/$ by formula (\ref{eq:norm}). Then
it is clear that axiom (A1) holds.  Given two fields $\phi (z)\/$
and $\psi(z)\/$, if $[T, \phi (z) ] = \partial_z \phi (z)\/$ and
$[T, \psi (z) ] = \partial_z \psi (z)\/$, then from
(\ref{eq:normal}) and (\ref{eq:rule}) it follows that
\begin{equation}
  [T, :\phi (z) \psi (z) :] = \partial_z : \phi (z)\psi (z) :.
\label{eq:ind}
\end{equation}
Hence the axiom $(A2)\/$ follows inductively from (\ref{eq:norm}) and
(\ref{eq:ind}).

Using an argument of Dong ({\em cf.} [L, Proposition 3.2.7]), one can show
that if three fields $\chi (z)\/$, $\phi (z)\/$ and $\psi(z)\/$ are
ultralocal with respect to each other, then $:\phi (z)\psi(z)\/$: and $\chi
(z)\/$ are ultralocal. It is also clear that if $\phi (z)\/$ and $\psi
(z)\/$ are ultralocal, then $\partial_z \phi (z)\/$ and $\psi (z)\/$ are
ultralocal. This implies axiom (A3).

Finally, from the uniqueness theorem of [G] it follows that the map $Y
(\cdot, z)\/$ is independent of the choice of the basis.\qed

We shall say that the VOA constructed in Proposition
\ref{prop:generator} is {\em generated\/} by the fields
$Y(a^{(i)}, z), i > 0\/$.

Now fix $c \in \Bbb C\/$, and consider the vacuum $\hd\/$-module
$M_c = M_c (\widehat{\cal D},{\cal P})\/$.  The space $M_c\/$ is
$\Bbb Z_{+}\/$-graded by eigenspaces of the operator $-D\/$: $M_c
= \oplus_{j \in \Bbb Z_{+}} M_{c, j}\/$, so that $M_{c,0} = \Bbb
C |0\rangle\/$, where $|0\rangle = 1\otimes 1\/$.  Recall that
$J_k^l |0\rangle = 0\/$ for $l \in \Bbb Z_{+}\/$ and
$k+l \geq 0\/$. Note that vectors of the form
$$
J^{l_n}_{-k_n -l_n -1} \ldots J^{l_1}_{-k_1 -l_1 -1}|0\rangle,
$$
where $(l_i , k_i) \in \Bbb Z_{+}^2\/$ span $M_c\/$. It follows
that the generating fields $J^l (z) = \sum_{k \in \Bbb Z} $
\linebreak
$J^l_k z^{-k-l-1}\/$ satisfy conditions (2) and (3) of Proposition
\ref{prop:generator}.  Condition (4) clearly holds.  Condition
(1) follows from the operator product expansion [R]:
\begin{equation}
  \everymath{\displaystyle}
  \begin{array}{rcl}
    J^m (z) J^n (w) & \sim &
    \sum^{m+n}_{a = 1} \left( [n]_a J^{m+n-a} (w) - (-1)^a [m]_a
    J^{m+n-a} (z) \right) / (z-w)^{a+1}\\
    && \quad +(-1)^m  m!n! c/(z-w)^{m+n+2}.
  \end{array}
\label{eq:ope}
\end{equation}
Hence the vacuum $\hd\/$-module $M_c\/$ is a VOA. It follows (by
skewsymmetry of vertex operators) that any quotient of the
$\hd\/$-module $M_c\/$ is a VOA. In particular, the irreducible
vacuum module $V_c = V_c (\hd, 0 )\/$ is a VOA.

Let now $\omega (\beta) = (J^1_{-2} + \beta J^0_{-2}) |0\rangle
\/$. It is easy to see that $Y(\omega (\beta), z) = \sum_{k \in
{\Bbb Z}} L_k^{+}(\beta)$ \linebreak
$z^{-k-2}\/$, where $L_k^{+}(\beta)\/$ is
defined in (\ref{eq:vira}). We know from (\ref{eq:family}) that
$L_k^{+}(\beta)\/$ generate the Virasoro algebra $Vir(\beta)\/$
with central element $C_{\beta}\/$. Furthermore from
(\ref{eq:bracket}) and (\ref{eq:vira}) it is easy to see that the
axiom (C) of the Virasoro element holds.  Thus, we have
established the following result:

\begin{theorem} \label{th:2}
  For any $c$ and $\beta$, the quadruple $\left( M_c ,|0\rangle,
  \omega(\beta), Y(\cdot,z) \right)$ is a conformal VOA of rank
  $c_{\beta} = -(12 \beta^2 -12 \beta +2)c$ generated by the
  fields $J^l (z)$, $l =0,1,2, \ldots$, of conformal dimension
  $l+1$. The same holds for $V_c$.
\end{theorem}

\begin{remark}
  Recall that a field $\phi(z)\/$ is called {\em primary\/} of
  conformal dimension $\Delta\/$ with respect to a Virasoro
  element $\omega\/$, if the following operator product expansion
  holds:
  $$
  Y (\omega, z) \phi(w) \sim \frac{\Delta \phi(w)}{(z-w)^2} +
  \frac{\partial_w \phi(w)}{z-w}.
  $$
  The field $J^0 (z)\/$ is primary with respect to $Y (w (\beta),
  z)\/$ if and only if $\beta = \frac12\/$ (in this case the rank
  of $M_c\/$ equals $c\/$).  For $l > 0\/$ one can always add to
  the field $J^l(z)\/$, $l \geq 0\/$, a normally ordered
  combination of the fields $J^k (z)\/$, $0 \leq k < l$, so that
  the resulting field is primary of conformal dimension $l+1\/$
  with respect to $\omega(\beta)\/$ for all but a finite number
  of values of $\beta\/$.
\label{rem:1}
\end{remark}

\section{Vertex Operator Algebra of $\W_{\beta}(gl_N)\/$}
\setcounter{equation}{0}

Let $\frak g\/$ be the Lie algebra $gl_N (\Bbb C )\/$ or $sl_N
(\Bbb C )\/$.  Let $\frak h\/$ be the corresponding subalgebra of
diagonal matrices. We denote by $\Delta \subset {\frak h}^*\/$ be
the set of roots, $\Delta_{+}\/$ the set of positive roots
corresponding to upper triangular matrices, and let $\alpha_1 ,
\ldots , \alpha_{N-1} \/$ be the simple roots. We identify
${\frak h}^*\/$ with ${\frak h}\/$ using the trace form $(a,b) =
\tr\,\, ab\/$ on $gl_N\/$, so that $(\alpha, \alpha) = 2\/$ for
$\alpha \in \Delta\/$.

Let $\widehat{\frak g} = \frak g \otimes \Bbb C [t, t^{-1}]
\oplus \Bbb C \bf k \/$ be the affine algebra associated to
$(\frak g, (\cdot,\cdot ))\/$ and let $\widehat{\frak h} = \frak
h \otimes \Bbb C [t,t^{-1}] \oplus \Bbb C \bf k\/$ be the
homogeneous Heisenberg subalgebra of $\widehat{\frak g}\/$.  The Lie
algebra $\widehat{\frak h}\/$ has generators $u(n), u \in \frak h, n
\in \Bbb{Z}\/$ with commutation relations
$$
  [u(m), v(n)] = m \delta_{m, -n} (u,v) {\bf k} .
$$
Given $\gamma \in \frak h^{*}\/$, denote by $\pi_{\gamma} =
\pi_{\gamma} (\frak h)\/$ the space of the irreducible
representation of the Lie algebra $\widehat{\frak h}\/$ which admits
a non-zero vector $|\gamma\rangle\/$ such that
$$
u(n) |\gamma\rangle = \delta_{n,0}
\gamma (u) |\gamma\rangle, \;
\mbox{ for } n \geq 0, \quad
{\bf k} |\gamma\rangle = |\gamma\rangle.
$$

By Proposition 3.1, the space ${\pi}_0 \/$ has a structure of a
VOA generated by the fields $u(z) = \sum_{n \in \Bbb Z} u(n)
z^{-n-1}\/$, $u \in \frak h\/$.  This VOA has the well-known
family of conformal structures given by the Virasoro element (in
the $sl_N\/$ case the linear term should be dropped):
$$
\omega_a = \hf \sum_i u_i (-1) u^i (-1) |0\rangle + aI(-2) |0
\rangle, \quad (a \in \Bbb C),
$$
where $\{u_i \}\/$ and $\{u^i\}\/$ are dual bases of $\frak h\/$
and $I\/$ is the identity matrix in $gl_N\/$. The central charge
of the corresponding Virasoro field $Y( \omega_a, z)\/$ for
$sl_N\/$ (resp.\ $gl_N\/$) is $N-1\/$ (resp.\ $N (1 - 12
a^2)\/$). The corresponding $\Bbb Z_{+}\/$-gradation ${\pi}_0 =
\oplus_{j \in \Bbb Z_{+}} \pi_{0,j}\/$ is given by $\wt |0\rangle
= 0\/$, $\wt\, u(n) = -n\/$.

There exists a unique operator $e^\gamma : \pi_0 \rightarrow
\pi_{\gamma}\/$ which maps $|0\rangle \/$ to $|\gamma\rangle \/$
and which commutes with all operators $u(n)\/$ with $u \in \frak
h, n \neq 0\/$. Let
$$
X_{\gamma} (z) = e^{\gamma} \exp \left( - \sum_{n<0}
\frac{\gamma(n)z^{-n}}{n} \right) \exp \left( -\sum_{n>0}
\frac{\gamma(n)z^{-n}}{n} \right),
$$
and let $X_{\gamma} (z) = \sum X_{\gamma} (n) z^{-n}\/$ be its
Fourier expansion where $X_{\gamma} (n)\/$ are linear operators
from $\pi_0 \/$ to $ \pi_{\gamma}\/$.

Given $\beta \in \Bbb C\/$, let
$$
\overline{\W}_{\beta}( \frak g) = \bigcap_{i=1}^{N-1}
\mbox{Ker}_{\pi_0} X_{\beta \alpha_i}(1).
$$
This is a vertex operator subalgebra of the VOA ${\pi_0}\/$ ({\em
cf.} [FF2, Lemma 4.2.8]).

The VOA $\overline{\W}_{\beta}( \frak g) \/$ is a $\Bbb
Z_{+}\/$-graded subspace of $\pi_0\/$, {\em i.e.},
$\overline{\W}_{\beta}( \frak g) = \oplus_{j \in \Bbb Z_{+}}
\overline{\W}_{\beta}( \frak g)_j\/$, where
$\overline{\W}_{\beta}( \frak g)_j\/$ is a subspace of the
(finite-dimensional) vector space $\pi_{0,j}\/$. It is clear that, given
$j\/$, for all but finitely many $\beta \in \Bbb C\/$ the dimension of
$\overline{\W}_{\beta}( \frak g)_j\/$ is the same (say $a_j\/$) and is
minimal.  Such $\beta\/$ is called {\em $j$-generic\/}. The value
$\beta \in \Bbb C\/$ which is $j$-generic for all $j \in \Bbb Z_{+}\/$ is
called {\em generic}. Thus for each $j\/$ we have a rational map of $\Bbb
C\/$ in the Grassmannian of $a_j\/$-dimensional subspaces in $\pi_{0, j}\/$
given by $\beta \mapsto \overline{\W}_\beta (\frak g)_j\/$.  This allows
us to define for an arbitrary $\beta_0
\in \Bbb C\/$, $\beta_0 \neq 0\/$, the analytic continuation:
$\W_{\beta_0} (\frak g) = \oplus_{j \in \Bbb Z_{+}}
{\W}_{\beta_0} (\frak g)_j\/$, where ${\W}_{\beta_0}( \frak g)_j
= \lim_{\beta \rightarrow \beta_0} \overline{\W}_{\beta}(\frak
g)_j\/$ and the limit is taken over the set of generic $\beta\/$.
Thus $\W_{\beta} (\frak g)\/$ is a family of vertex operator
subalgebras of $\pi_0\/$, which depends on $\beta\/$. This family
of VOA is called the family of $\W\/$--algebras of $\frak g\/$.

\begin{remark}
\label{rem:2}
  On a more formal level, consider $\beta\/$ as a formal
  variable, {\em i.e.}, consider $\pi_0\/$ and $\pi_{\beta
    \alpha_i}\/$ as free modules over the ring ${\Bbb C}
  [\beta]\/$.  Then the intersection of the kernels of
  operators $X_{\beta \alpha_i}(1): \pi_0 \rightarrow \pi_{\beta
    \alpha_i}\/$, $i=1,\ldots,l\/$, is also a free $\Bbb C
  [\beta]\/$-module. For any $\beta_0 \neq 0\/$,
  $\W_{\beta_0} (\frak g)\/$ is then defined by the
  specialization of this ring at $\beta = \beta_0\/$, {\em i.e.},
  as the quotient of this kernel by the submodule generated by
  $(\beta-\beta_0)\/$. Clearly, $\W_{\beta_0} (\frak g)\/$ is a
  vertex operator subalgebra of $\overline{\W}_{\beta} (\frak
  g)\/$, and it coincides with $\overline{\W}_{\beta}( \frak
  g)\/$ for generic $\beta_0\/$.
\end{remark}

\begin{remark}
  The VOA ${\W}_{\beta}(gl_N )\/$ (resp.\
  $\overline{\W}_{\beta} (gl_N)\/$) is isomorphic to the tensor
  product of the VOA ${\W}_{\beta}(sl_N )\/$ (resp.\
  $\overline{\W}_{\beta}(sl_N )\/$) and the VOA associated to a
  free bosonic field (of conformal dimension 1).

  The VOA ${\W}_{\beta}(sl_N )\/$ coinsides with the
  $\W\/$--algebra defined in [Z] for $N=3\/$ and in [FL] for
  general $N\/$.
\label{rem:ww}
\end{remark}

The following is a corollary of Theorem 4.5.9 from [FF2].

\begin{theorem}    \label{th:free}
  The VOA $\W_{\beta}(gl_N)\/$ is freely generated by
  fields $W^{(i) ,\beta} (z)\/$, $i=0,1,
  \ldots, N-1\/$, of conformal dimensions $i+1\/$.
\end{theorem}

\begin{remark}
  Note that $\omega_a \in \W_1 (gl_N)\/$; recall that the central charge is
  equal to $N (1 - 12 a^2)\/$. The following theorem was conjectured in
  [Bo] and its proof was indicated in [BS] ({\em cf.} also [BBSS]).
\label{rem:3}
\end{remark}

\begin{theorem} \label{th:gener}
  $\W_1 (\frak g) = \overline{\W}_1 (\frak g)\/$.
\end{theorem}

{\em Proof.} Due to Remark \ref{rem:ww}, it suffices to prove the
theorem for $\frak g =sl_N\/$. By definition, $\W_{\beta} (sl_N)
\subset \overline{\W}_{\beta} (sl_N)\/$, for any $\beta\/$.  The
theorem now follows from the comparison of characters of $\W_1
(sl_N)\/$ and $\overline{\W}_1 (sl_N)\/$.

It follows from Theorem \ref{th:free} that for generic $\beta\/$,
$\overline{\W}_{\beta}\/$ has a basis, which consists of
lexicographically ordered monomials in the following Fourier
components of the fields $W^{(i),\beta}(z) = \sum_{n \in \Bbb Z}
W^{(i),\beta}_n z^{-n-i-1}: \left\{ W^{(i),\beta}_n, n \leq
{-i-1}, i=1, \ldots, N-1 \right\}\/$. Therefore we have for a
generic $\beta\/$:
\begin{equation}
  \ch \W_1(sl_N) = \ch \W_{\beta}(sl_N) = \ch \overline{\W}_{\beta}
  (sl_N) = \prod_{i=1}^{N-1} \prod_{j=1}^{\infty} (1-q^{i+j})^{-1}.
\label{eq:equality}
\end{equation}
On the other hand, due to the vertex operator construction [FK]
of the basic $\widehat{\frak g}\/$-module $L(\Lambda_0)\/$, we
have:
\begin{equation}
  \pi_0 = \left\{ v \in L(\Lambda_0)| \frak h \cdot v = 0 \right\},
\label{eq:invt}
\end{equation}
\begin{equation}
  E_{\alpha_i} (0) \mid_{\pi_0} = X_{\alpha_i} (1).
\label{eq:fourier}
\end{equation}
By the complete reducibility of the $\frak g\/$-module
$L(\Lambda_0)\/$, we conclude from (\ref{eq:invt}) and
(\ref{eq:fourier}) that
\begin{equation}
  \overline{\W}_1 (sl_N) = \left\{ v \in L(\Lambda_0)|\frak g
  \cdot v = 0 \right\}.
\label{eq:coins}
\end{equation}
But the character of the right-hand side of (\ref{eq:coins}) is
known [K1, Proposition 2], which gives us
\begin{equation}
  \everymath{\displaystyle}
  \begin{array}{rcl}
    \ch \overline{\W}_1(sl_N)
    &=& \prod_{\alpha \in \Delta_{+}}
    \left( 1-q^{({\rho}, \alpha)} \right)/ \varphi (q)^{N-1} \\
    &=&\prod_{1\leq i \leq j \leq N-1} \left(1-q^{j-i+1}\right)/
    \varphi (q)^{N-1} \\
    &=&\prod_{i=1}^{N-1} \prod_{j=1}^{\infty}
    \left(1-q^{i+j}\right)^{-1},
  \end{array}
\label{eq:charinvt}
\end{equation}
where $\rho = \frac{1}{2} \sum_{\alpha \in \Delta_{+}} \alpha\/$.
Comparing (\ref{eq:equality}) and
(\ref{eq:charinvt}) completes the proof.  \qed

\begin{corollary} \label{cor:1}
Let $\frak g = gl_N\/$. Then the intersection of the kernels of the
operators $X_{\alpha_i}(1)\/$ on ${\pi}_0\/$ coincides with $\W_1
(gl_N)\/$. Furthermore one has:
\begin{equation}
  \ch \W_1(gl_N) = \prod_{i=0}^{N-1}
  \prod_{j=1}^{\infty} \left( 1-q^{i+j} \right)^{-1}.
\label{eq:12}
\end{equation}
\end{corollary}

\begin{remark}
  As in [FF2], $\overline{\W}_{\beta}( \frak g)\/$ and ${\W}_{\beta}( \frak
  g)\/$ may be defined for an arbitrary simple Lie algebra in the same way
  as for $\frak g=sl_N\/$.  The same argument as above shows that for an
  arbitrary simply-laced simple Lie algebra $\frak g\/$, $\W_1(\frak g) =
  \overline{\W}_1(\frak g)\/$, {\em cf.} also [Bo, BS] (note that this is
  not true for non-simply laced $\frak g$).

The Fourier components of vertex operators from the VOA $\pi_0$ span a Lie
algebra $U(\widehat{\frak h})_{loc}$, which lies in a completion of
$U(\widehat{\frak h})/({\bf k}-1)U(\widehat{\frak h})$ [FF1]. The Fourier
components of vertex operators from the VOA $\W_\beta(\frak g)\/$ span a
Lie subalgebra of $U(\widehat{\frak h})_{loc}$, which we denote by
$U\W_\beta(\frak g)_{loc}\/$. This Lie algebra is also called
$\W$--algebra of $\frak g$.
\label{rem:local}
\end{remark}

\begin{remark}
The Lie algebra $U\W_1(\frak g)_{loc}\/$ (for simply-laced $\frak g$)
was considered for the first time in [F2]. More precisely, in [F2] the Lie
algebra $\widehat{S}^W$ was defined, which is linearly spanned by all
Fourier components of vertex operators from $\pi_0$, which commute with the
action of $\frak g$ and which are invariant with respect to the natural
action of the Weyl group $W$ of $\frak g$ on $\pi_0$.

The Lie algebra $\widehat{S}^W$ coincides with $U\W_1(\frak g)_{loc}\/$.
Indeed, according to Theorem~\ref{th:gener}, the Lie algebra $U\W_1(\frak
g)_{loc}\/$ consists of all Fourier components of vertex operators from
$\pi_0$, which commute with the operators $X_{\alpha_i}(1),
i=1,\ldots,N-1$. This is equivalent to commuting with $\frak g$. On the
other hand, it is easy to show that all elements of $U\W_1(\frak
g)_{loc}\/$ are automatically $W$--invariant.

This is obvious in the case ${\frak g}=sl_2$, because $U\W_1(sl_2)_{loc}\/$
is generated by the Fourier components of the vertex operator $\frac{1}{2}
:u(z)^2:$, which is invariant under the transformation $u(z) \rightarrow
-u(z)$. This fact and Theorem~\ref{th:gener} imply that all elements of
$U\W_1(\frak g)_{loc}\/$ for an arbitrary simply-laced ${\frak g}$ are
invariant under the simple reflections from $W$, and hence are
$W$--invariant (note however that not all $W$--invariant elements of
$U(\frak h)_{loc}$ belong to $U\W_1(\frak g)_{loc}$).

Moreover, it follows from Theorem~4.5.9 of [FF2] and Theorem~\ref{th:gener}
that if $\frak g$ is simply-laced, and $P_1({\bf u}),\ldots,P_l({\bf u})$
is a set of generators of the ring ${\Bbb C}[\frak h]^W$, then the fields
$:P_1({\bf u}(z)):,\ldots,:P_l({\bf u}(z)):$ freely generate the VOA
$\W_1(\frak g)$.
\label{rem:4}
\end{remark}

\section{Connection between $\hd\/$ and $\W_1 (gl_N)\/$}
\setcounter{equation}{0}

Due to Proposition \ref{prop:generator}, $\cal F\/$ is a (super) VOA
with the generating (odd) fields
$$
\psi (z) = \sum_{i \in \Bbb Z} \psi_i z^{-i-1},\quad
\psi^* (z) = \sum_{i \in \Bbb Z} \psi_i^{*} z^{-i}.
$$
Recall that $\alpha (z) = :\psi (z) \psi^{*} (z): = \sum_{n \in
\Bbb Z} \alpha_n z^{-n-1}\/$ is a free bosonic field, {\em i.e.},
$[\alpha_m, \alpha_n ] = m \delta_{m, -n}\/$, and that each
${\cal F}^{(m)}\/$, the subspace of $\cal F\/$ of charge $m\/$,
is irreducible with respect to the $\alpha_n\/$ (this follows
from formula (\ref{eq:char}) for $c=1\/$). It follows that ${\cal
F}^{(0)}\/$ is a vertex operator subalgebra isomorphic to the VOA
$\pi_0 (\Bbb C)\/$.

Note that the homomorphism $\widehat{\phi}_0 : \hd \rightarrow
\hgl\/$ induces a $\hd\/$-module homomorphism
$$
\epsilon: M_1 = U (\hd) \otimes_{U (\widehat{\cal P})}
\Bbb C_0 \longrightarrow {\cal F}^{(0)}.
$$
Similarly, the homomorphism $\widehat{\phi}_0^N: \hd
\longrightarrow \hgl^N\/$ induces the homomorphism of
$\hd\/$-modules
$$
\epsilon_N : M_N \rightarrow \left({\cal F}^{(0)}\right)^{\otimes N}.
$$
On the level of operators the map $\epsilon_N \/$ is given by
\begin{equation}
  J^l (z) \mapsto \sum^N_{i=1} :\psi^i (z) \partial_z^l \psi^{*i}
  (z):,
\label{eq:realize}
\end{equation}
where $\psi^i (z)\/$ and $\psi^{*i} (z)\/$ denote the fermionic
fields on the $i\/$-th copy of $\cal F\/$. Recall also the
following formula of the well-known boson-fermion correspondence
(see {\em e.g.}, [K2, Chapter 14])
\begin{equation}
  X_{\alpha_i} (z) = :\psi^i (z) \psi^{*i+1}(z):.
\label{eq:boson}
\end{equation}

\begin{lemma} \label{lem:incl}
  $\mbox{\rm Im}\, \epsilon_N \subset \W_1(gl_N)\/$.
\end{lemma}

{\em Proof.} Due to (\ref{eq:boson}), $\W_1(gl_N)\/$ is the
intersection of the kernels of operators $A_i = \int :\psi^{i}
(z) \psi^{* i+1}(z):\, dz\/$.  Since $|0\rangle \in
\widehat{\pi}_0\/$ is in the kernels of these operators, and
$M_N\/$ is generated by the Fourier components of $J^l (z), l
\geq 0\/$, applied to $|0\rangle\/$, it suffices to check that
$J^l (z)\/$ commutes with $A_i\/$. Due to (\ref{eq:realize}), we
have to check that
$$
\left[
  \sum_{k=1}^{N} : \psi^k (z) \partial^m \psi^{*k} (z):,
  \int : \psi^i (u) \psi^{* i+1}(u): \, du
\right] = 0
$$
for any $m\geq 0\/$ and $i= 1, \ldots, N-1\/$.  In fact, a much
stronger statement holds:
\begin{equation}
  \left[
    \sum^N_{k=1} : \psi^k (z) \psi^{*k} (w) :,
    \int : \psi^i (u) \psi^{*j} (u):\, du
  \right] = 0.
  \label{eq:zero}
\end{equation}
In order to prove (\ref{eq:zero}) we calculate the operator
product expansions (OPE).  We have
$$
\psi^m (z)\psi^{*n}(w) \sim \frac{\delta_{m,n}}{z-w}, \quad
\psi^{*m}(z) \psi^n (w)\sim \frac{\delta_{m,n}}{z-w}.
$$
By the Wick theorem, we have
\begin{displaymath}
  \left(
    \sum_{k=1}^N : \psi^k (z) \psi^{*k} (w) :
  \right)
  \left(
    : \psi^i (u) \psi^{*j} (u) :
  \right) \quad \sim \quad
  \frac{:\psi^i (z) \psi^{*j} (u):}{w - u} +
  \frac{:\psi^{*j} (w) \psi^i (u):}{z - u}.
\end{displaymath}
But for local fields $a (z)\/$ and $b (z)\/$ with OPE $a (z) b(u)
\sim \sum_j c_j (z) / (z - u)^j\/$ we have $\left[a (z), \int b
(u)\, d u \right] = c_1 (z)\/$.  Hence the left-hand side of
(\ref{eq:zero}) is equal to $:\psi^i (z) \psi^{*j} (w): + :\psi^{*j}
(w) \psi^i (z): = 0\/$. \qed

By Lemma \ref{lem:incl}, we have a $\hd\/$-module homomorphism
$\epsilon_N : M_N \rightarrow \W_1(gl_N)\/$. We know that (see
(\ref{eq:equality})):
$$
\ch {\W}_1(gl_N) =\prod_{i=0}^{N-1}  \prod_{j=1}^{\infty}
(1-q^{i+j})^{-1} \equiv \sum_{n\geq 0} a_n q^n,
$$
and clearly
$$
\ch M_N =\prod_{k=1}^{\infty} \left( 1 - q^k \right)^{-k}
\equiv \sum_{n \geq 0} b_n q^n.
$$

Observe that $\ch M_N \/$ and $\ch {\W}_1(gl_N)\/$ coincide from
weight 0 to weight $N\/$, {\em i.e.}, $a_n = b_n\/$ for $ n =
0,\ldots,N\/$ and, moreover, $a_{N+1} = b_{N+1} +1\/$. It follows
that the lowest nonzero weight of the kernel of $ \epsilon_N \/$
is $N+1\/$. From [KR, Example 5.2] we know that there exists a
unique singular vector $v\/$ of weight $N+1\/$ such that the
submodule $\langle v \rangle \/$ generated by $v\/$ is the
maximal proper submodule in $M_N\/$, {\em i.e.}, $V_N = M_N /
\langle v \rangle \/$ is irreducible. Hence we see that the
kernel of $ \epsilon_N \/$ is nothing but the submodule $\langle
v \rangle \/$. The homomorphism $\epsilon_N : M_N \rightarrow
\W_1(gl_N)\/$ therefore induces an injective $\hd\/$-module
homomorphism $\eta_N : V_N \rightarrow \W_1(gl_N)\/$.  By
comparing specialized characters (see (\ref{eq:char})), we have
$\ch V_N = \ch {\W}_1(gl_N)\/$. Thus we have proved the following
fact.

\begin{theorem} \label{th:7}
The map $\epsilon_N : M_N \longrightarrow \W_1(gl_N)\/$ induces a
$\hd\/$-module isomorphism $ \eta_N : V_N
\stackrel{\sim}{\rightarrow} \W_1(gl_N)\/$, which is also an
isomorphism of VOAs.  One has:
\begin{displaymath}
  \epsilon_N \left(\omega (\beta)\right) = \omega_{\beta - 1/2}.
\end{displaymath}
\end{theorem}

\begin{remark} To any VOA $V$ one can canonically associate a Lie algebra,
which consists of all Fourier components of vertex operators from $V$, {\em
cf.}  [FF1]. For the VOA $M_N$ this Lie algebra, which we denote by
$U_N(\hd)_{loc}$, lies in a certain topological completion of
$U(\hd)/(C-N)U(\hd)$. We call $U_N(\hd)_{loc}$ the local completion of
$U(\hd)$. Denote by $s_n, n \in {\Bbb Z}$, the Fourier components of the
vertex operator $Y(S,z)$, where $S$ is a singular vector of degree $N+1$ in
$M_N$. Now let $U\W_1(gl_N)_{loc}$ be the Lie algebra of all Fourier
components of vertex operators from the VOA $\W_1(gl_N)$, which was defined
in Remark~\ref{rem:local}. By Theorem~\ref{th:7}, $U\W_1(gl_N)_{loc}$ is
the quotient of the Lie algebra $U_N(\hd)_{loc}$ by the ideal generated by
$s_n, n \in {\Bbb Z}$.
\end{remark}

\begin{corollary} \label{cor:2}
  $\W_1(gl_N)\/$ is a simple VOA.
\end{corollary}

{\em Proof.} Any ideal of ${\W}_1(gl_N)\/$ as VOA can be regarded
as a $\hd\/$-module via $\epsilon_N \/$. Since ${\W}_1(gl_N)
\cong V_N \/$ as a $\hd\/$-module is irreducible, there are no
nontrivial ideals of ${\W}_1(gl_N)\/$ as VOA. \qed

\begin{corollary}    \label{cor:lift}
Any representation of the VOA $\W_1(gl_N)\/$ can be canonically
lifted to a representation of the Lie algebra $\widehat{\cal
  D}\/$ with central charge $N\/$.
\end{corollary}

Let $\frak h\/$ be the Cartan subalgebra of $gl_N\/$ (as in
Sect.~4). For $\gamma \in \frak h^*\/$ let $\gamma_i =
\gamma(E_{ii})\/$, $i = 1, \ldots, N\/$. Recall that each
$\pi_{\gamma}\/$ ($\gamma \in \frak h^*\/$) is a
representation space of the VOA $\pi_0\/$, hence of the VOA $\W_1
(gl_N)\/$.  Denote by $V_N(\gamma)\/$ the irreducible quotient of
the $\W_1(gl_N)\/$-submodule of $\widehat{\pi}_{\gamma}\/$
generated by the highest weight vector $|\gamma \rangle \/$.

\begin{proposition}    \label{prop:hw}
  (a) The modules $V_N(\gamma)\/$ are all up to isomorphism
  irreducible modules over the VOA $\W_1(gl_N)\/$.

  (b) The lifting of a module $V_N(\gamma)\/$ to $\hd\/$ is
  isomorphic to the primitive $\hd\/$-module with exponents
  $\gamma_1, \ldots, \gamma_N\/$.
\end{proposition}

{\em Proof.} (a) follows from the fact that irreducible
$\W_1(gl_N)\/$-modules are determined by the highest weights. In
order to prove (b) note that $V_N (\gamma)\/$ is an irreducible
highest weight $\cal D\/$-module with $c = N\/$ and one finds by
a direct computation that $\Delta_\lambda (x) = \sum^N_{i=1}
\frac{e^{\gamma_i x} - 1}{e^x - 1}\/$.
\qed

Thus, the primitive $\hd\/$-modules with central charge $N\/$
produce all irreducible $\W_1(gl_N)\/$-modules. In particular,
from Theorem \ref{th:sumch} and formula (\ref{eq:char}) one obtains the
complete and specialized characters of all irreducible
$\W_1(gl_N)\/$-modules.  The $W_1 (gl_N)\/$-modules $V_N
(\gamma)\/$ with integral $\gamma\/$ were considered in [BMP],
where they were used in the study of semi-infinite cohomology of
$W\/$-gravity models.

\begin{remark} \label{rem:5}
  Fix $r \in \Bbb C\/$ and consider the associative algebra
  $Cl_r\/$ on generators $\psi_j\/$ ($j \in - r + \Bbb
  Z\/$) and $\psi^*_j\/$ ($j \in r + \Bbb Z\/$) with defining
  relations (\ref{eq:9}), and let $\psi (z) = \sum_{j \in -r +
    \Bbb Z} \psi_j z^{-j -r -1}\/$, $\psi^* (z) = \sum_{j \in r +
    \Bbb Z} \psi^*_j z^{-j + r}\/$.  Let ${\cal F}_r\/$ denote
  the unique irreducible $Cl\/$-module such that
  \begin{displaymath}
    \psi_j | r \rangle = 0 \mbox{ if } j + 1 + r > 0, \quad
    \psi^*_j | r \rangle = 0 \mbox{ if } j - r > 0.
  \end{displaymath}
  Then formula (\ref{eq:realize}) gives a primitive
  representation of $\widehat{\cal D}\/$ in ${\cal F}^{(0)}_r\/$,
  with central charge 1 and exponent $r\/$.
\end{remark}

\begin{remark} \label{rem:7}
  Summarizing, any positive energy $\widehat{\cal D}\/$-module
  $M\/$ gives rise to a module over the associated VOA $M_c\/$.
  If $c \notin \Bbb Z\/$, then $M_c = V_c\/$.  If $c \in \Bbb
  Z_+\/$, then any primitive $\widehat{\cal D}\/$-module gives
  rise to a module over the VOA $V_c\/$ and all irreducible
  modules over $V_c\/$ are thus obtained.
\end{remark}

\begin{remark} \label{rem:6}
  In [Zh], Zhu constructed an associative algebra $A(V)\/$
  corresponding to an arbitrary VOA $V\/$ and established a one
  to one correspondence between irreducible representations of
  $V\/$ and irreducible representations of $A(V)\/$.  In our
  case, one can show that the associative algebra $A(M_c)\/$, $c
  \in \Bbb C\/$, is isomorphic to the polynomial algebra in
  infinitely many variables $w_0, w_1, \ldots\/$, which
  correspond to the generating fields $J^0(z), J^1 (z),
  \ldots\/$, and that $A(V_N)\/$ for $N \in \Bbb Z_+\/$ is
  isomorphic to the polynomial algebra $\Bbb C [ w_0, w_1, \ldots
  , w_{N-1}]\/$.
\end{remark}

\section{Towards fusion rules for $\widehat{\cal D}$}
\setcounter{equation}{0}

We can define spaces of conformal blocks for the Lie algebra $\hd$ in the
same fashion as for the Virasoro or affine algebras, using coinvariants,
cf., e.g., [FFu].

For simplicity we will restrict ourselves to the genus 0 case.  Consider
the projective line ${\Bbb C}{\Bbb P}^1$ with a global coordinate $t$ and
$n$ marked points: $z_1,\ldots,z_n$. We assume that $z_i \neq
\infty$ for all $i=1,\ldots,n$. Around the point $z_i$ we have a local
coordinate $t-z_i$. Denote by ${\cal D}(z_i)$ the Lie algebra of
differential operators on the formal punctured disc around $z_i$. Elements
of this Lie algebra are finite sums $$\sum_{m \in \Bbb Z_{+}} f_m(t-z_i)
\, (\partial_t)^m, \quad \mbox{where  } f_m(t-z_i) \in {\Bbb C}((t-z_i)).$$

Now let $\hd(z_1,\ldots,z_n)$ be the central extension by $\Bbb C$ of
the Lie algebra $\oplus_{i=1}^n {\cal D}(z_i)$, such that its restriction
to each of the summands ${\cal D}(z_i)$ coincides with the one
defined by the 2-cocycle (\ref{eq:cocy}).

Denote by ${\cal D}_{z_1,\ldots,z_n}$ the Lie algebra of regular
differential operators on ${\Bbb C}{\Bbb P}^1 \backslash \{ z_1,\ldots,$
$z_n \}$. We have a natural embedding of this Lie algebra into the Lie
algebra $\oplus_{i=1}^n {\cal D}(z_i)$, obtained by expanding a
differential operator around each of the points $z_i$. One can check that
the restriction of the 2-cocycle to ${\cal D}_{z_1,\ldots,z_n}$ is trivial,
and therefore we obtain an embedding ${\cal D}_{z_1,\ldots,z_n}
\rightarrow \hd(z_1,\ldots,z_n)$.

Let $M_1,\ldots,M_n$ be highest weight $\hd$--modules with the same central
charge. Then the tensor product $M_1 \otimes \ldots \otimes M_n$ is a
$\hd(z_1,\ldots,z_n)$--module. We define the space of conformal blocks,
corresponding to these modules, as the space of coinvariants of $M_1
\otimes \ldots \otimes M_n$ with respect to the Lie algebra ${\cal
D}_{z_1,\ldots,z_n}$. We denote this space by $H(M_1,\ldots,M_n)$. In
particular, the case $n=3$ corresponds to the so-called fusion rules, which
can also be defined via the intertwining operators introduced in [FHL], cf.
[W].

A primitive $\hd\/$--module with positive integral central
charge $N\/$ is called {\em integral\/} if all its exponents $r_1,
\ldots, r_N\/$ are integral; denote this module by $V ({\bf r})\/$
(recall that all these modules are unitary).  Let $P = \Bbb Z^N\/$ and $P^+
= \left\{ {\bf r} \in P | \,\,\, r_1 \geq \ldots \geq r_N
\right\}\/$.  Putting the exponents in a decreasing order, we see
that the integral primitive $\hd$--modules are parametrized
by $P^+: {\bf r} \mapsto V ({\bf r})\/$.  On the other hand, we may
view $P\/$ as the weight lattice of the group $GL_N (\Bbb C)\/$.
Then $P^+\/$ parametrizes the finite-dimensional rational
irreducible representation of $GL_N (\Bbb C) : {\bf r} \mapsto F
({\bf r})\/$, where $F ({\bf r})\/$ denotes the finite-dimensional
irreducible representation of $gl_N (\Bbb C)\/$ with highest
weight ${\bf r}\/$, ({\em i.e.\/}, ${\bf r} (E_{ii}) = r_i\/$).

\begin{conjecture}     \label{con:2}
The space $H(V({\bf r}_1),\ldots,V({\bf r}_n))$ is
isomorphic to the space of $gl_N({\Bbb C})$ -- invariants in the tensor
product $F({\bf r}_1) \otimes \ldots \otimes F({\bf r}_n)$.
In particular, for ${\bf r}, {\bf s} \in P^+\/$, let
  \begin{displaymath}
    F ({\bf r}) \otimes F ({\bf s}) = \bigoplus_{{\bf m} \in P^+}
    c^{\bf m}_{\bf rs} F ({\bf m})
  \end{displaymath}
  be the decomposition of the tensor product of $gl_N (\Bbb
  C)\/$-modules.  Then the fusion rules of primitive integral
  $\hd$-modules with central charge $N\/$ are given by the
  same formula:
  \begin{displaymath}
    V ({\bf r}) \cdot V ({\bf s}) = \bigoplus_{{\bf m} \in P^+}
    c^{\bf m}_{\bf rs} V ({\bf m}).
  \end{displaymath}
\end{conjecture}

\begin{example} \label{ex:one}
  (a) Since ${\cal W} (gl_1)\/$   is isomorphic to the VOA $\pi_0
  (\Bbb C)\/$, we have the fusion rules $V (r) \cdot V (s) = V
  (r+s)\/$.

  (b) ${\cal W} (gl_2)\/$ with $c = 2\/$ is isomorphic to the tensor
  product of the irreducible vacuum module with $c = 1\/$ over the Virasoro
  algebra and $\pi_0 (\Bbb C)\/$. Denote by $[n]$ the irreducible module
  over the Virasoro algebra with central charge $1$ and highest weight
  $n^2/4,\,n \in \Bbb Z_{+}$.
  Conjecture~\ref{con:2} states that the fusion rules for these
  modules are given by $$ [m] \cdot [n] = \sum_{k \in {\cal P}_{m,n}}
  [k],$$ where ${\cal P}_{m,n} = \{k |\/$ $|m-n|\leq k \leq m+n, m+n+k\/$
  is even$\}$, $m,n,k \in \Bbb Z_{+}\/$.
\end{example}

In order to provide some evidence for Conjecture~\ref{con:2}, let %
\begin{displaymath}
  M = \bigoplus_{\gamma \in P} \pi_\gamma ,
\end{displaymath}
where $\pi_\gamma\/$ is the irreducible $\widehat{\frak h}\/$-module
defined in Sect.~4, $\frak h\/$ being the Cartan subalgebra of
$\frak g = gl_N (\Bbb C)\/$.  Then the classical vertex operator
construction [FK] gives $M\/$ a structure of a unitary
$\widehat{\frak g}\/$-module of level 1.  More explicitly, $M\/$
decomposes into a direct sum of irreducible unitary $\widehat{\frak
g}\/$-submodules of level 1 with highest weight vectors $| \gamma
\rangle \in \pi_\gamma\/$, where $\gamma \in P\/$ are such that
$\gamma_i - \gamma_{i+1} = \delta_{is}\/$ for some $1 \leq s \leq N\/$
(here we put $\gamma_{N+1}=\gamma_1$), the corresponding $\widehat{\frak
g}\/$-submodule being $\bigoplus_{\alpha \in \gamma + Q} \pi_\alpha\/$,
where $Q = \left\{\beta \in P | \sum_i \beta_i = 0 \right\}\/$.

Viewed as a $\frak g\/$-module, $M\/$ decomposes into a direct
sum of finite-dimensional irreducible modules, which can be
integrated to $GL_N (\Bbb C)\/$.  On the other hand, each
$\pi_\gamma\/$ is a module over the VOA $\pi_0\/$, hence over the
VOA $\W_1 (gl_N)\/$.  Due to Theorem~\ref{th:7}, we see that each
$\pi_\gamma\/$ has a canonical structure of a $\hd$-module
with central charge $N\/$.  Thus, $M\/$ is a $\hd$-module
(and a $\W_1 (gl_N)\/$-module) with central charge $N\/$.
Moreover, it follows from the proof of Theorem~\ref{th:gener}
that the action of $\hd$ and $GL_N (\Bbb C)\/$ on $M\/$
commute.

\begin{theorem} \label{th:61}
  With respect to the commuting pair $\left(\widehat\D, GL_N (\Bbb C)
  \right)\/$ the module $M\/$ decomposes as follows:
  \begin{equation}
    M = \bigoplus_{{\bf r} \in P^+} V ({\bf r}) \otimes F ({\bf r}),
    \label{eq:61}
  \end{equation}
  the highest weight vector of $V ({\bf r}) \otimes F ({\bf r})\/$
  being $| {\bf r} \rangle\/$.
\end{theorem}

{\em Proof.} Recall that by the specialized character formula
(\ref{eq:char}) we have
\begin{equation}
  \tr_{V ({\bf r})} q^{L_0 (\frac{1}{2})} = q^{\frac 12 \sum_i r_i
  \left(r_i + 1 \right)} \prod_{1 \leq i < j \leq N}
  \left(1 - q^{r_i - r_j + j - i} \right) / \varphi (q)^N.
  \label{eq:62}
\end{equation}
On the other hand, denote by ${\cal U} ({\bf r})\/$ the direct
sum of all $GL_N (\Bbb C)\/$-submodules of $M\/$ isomorphic to $F
({\bf r})\/$.  Since $\hd$ commutes with $GL_N (\Bbb C)\/$,
this is a $\hd$-module.  It is clear that $| {\bf r} \rangle
\in V ({\bf r}) \subset {\cal U} ({\bf r})\/$ is the vector with
minimal eigenvalue of $L_0 \left(\frac 12\right)\/$ on ${\cal
U}({\bf r})\/$.  Comparing (\ref{eq:62}) with the character of
${\cal U} ({\bf r})\/$ computed in [K1] (see also [K2, Exercise
12.17]), we see that $\tr_{V ({\bf r})} q^{L_0 \left( \frac 12
\right)} = \tr_{{\cal U} ({\bf r})} q^{L_0 \left(\frac 12
\right)}\/$.  It follows that ${\cal U} ({\bf r}) = V ({\bf
r})\/$. \qed

Due to (\ref{eq:zero}) we have the following equivalent
formulation of Theorem~\ref{th:61}:

\begin{theorem} \label{th:62}
  The representation of $\widehat{gl}\/$ in ${\cal F}^{\otimes N}\/$ given
  by $\sum_{i, j\in \Bbb Z} E_{ij} z^{i - 1} w^{-j} \mapsto$
  \linebreak
  $\sum^N_{k=1} \psi^k (z) \psi^{*k} (w)\/$ and the representation
  of $gl_N (\Bbb C)\/$ in ${\cal F}^{\otimes N}\/$ given by
  \linebreak
  $E_{ij} \mapsto \int : \psi^i (z) \psi^{*j} (z) : dz\/$ ($i, j
  = 1, \ldots, N\/$) commute.  The decomposition of ${\cal
    F}^{\otimes N}\/$ with respect to the commuting pair
  $(\widehat{gl},gl_N (\Bbb C)\/)$ is as follows:
  \begin{equation}
    {\cal F}^{\otimes N} = \bigoplus_{{\bf r} \in P^+} V
    \left(
      \sum_i \widehat{\Lambda}_{r_i} \right) \bigotimes F ({\bf r}).
    \label{eq:63}
  \end{equation}
  (By restricting to $\widehat{\cal D}\/$ via the embedding
  $\phi_0\/$ this decomposition coincides with the decomposition
  (\ref{eq:61}) with respect to $(\widehat{\cal D},gl_N
  (\Bbb C)\/$).)
\end{theorem}

\begin{remark} \label{rem:61}
  The decomposition (\ref{eq:61}) is easy for $N = 1\/$. For $N = 2\/$ an
  equivalent form of (\ref{eq:61}) was established in [K]. The
  decomposition (\ref{eq:63}) for general $N$ was proved by another method
  in [F1, Theorem~1.6]; it also follows from [KP1, Proposition 1].
\end{remark}

\begin{remark} \label{rem:62}
  Another motivation of Conjecture \ref{con:2} is the fact that fusion
  rules given by this conjecture for $c = N$ coincide with the limit of the
  fusion rules for the $p$th unitary minimal model as $p$ goes to
  infinity.
\end{remark}

\frenchspacing
\noindent{\em Note added.} After this paper was finished, we saw on hep-th
net the paper by H. Awata, M. Fukuma, Y. Matsuo, and S. Odake ``Character
and determinant formulae of quasifinite representation of the
$\W_{1+\infty}$ algebra'' (hep-th/9405093), where character formulas for a
certain subclass of quasi-finite modules are given.

\end{document}